\documentclass{elsart5p}
\usepackage{graphics}
\usepackage{graphicx}
\usepackage{epsfig}
\usepackage{amssymb}
\usepackage{amsmath}
\begin{document}

\begin{frontmatter}

\title{Estimation of the ratio of the $pn{\to}pn\pi^0\pi^0/pn{\to}d\pi^0\pi^0$
cross sections}

\author[Uppsala]{G.~F\"aldt}\ead{goran.faldt@fysast.uu.se},
\author[UCL]{C.~Wilkin\corauthref{cor1}}
\ead{cw@hep.ucl.ac.uk}
\corauth[cor1]{Corresponding author.}

\address[Uppsala]{Department of Physics and Astronomy, Uppsala University,
Box 516, 751~20 Uppsala, Sweden}
\address[UCL]{Physics and Astronomy Department, UCL, London WC1E 6BT, UK}

\begin{abstract}
Evidence has recently been presented for the existence of a dibaryon
of mass 2380~MeV/$c^2$ and width 70~MeV/$c^2$, which decays strongly
into the $d\pi^0\pi^0$ channel [M.~Bashkanov \textit{et al.}, Phys.\
Rev.\ Lett.\ \textbf{102} (2009) 052301; P.~Adlarson \textit{et al.},
arXiv:1104.0123]. The decay rate of such a hypothesised dibaryon into
the $\{pn\}_{I=0}\pi^0\pi^0$ channel is estimated in a weakly
model-dependent way by using final state interaction theory. It is
shown that, if the resonance exists, it should then show up as
strongly in this channel as in $d\pi^0\pi^0$. The sum of the two
decay modes would saturate most of the inelasticity predicted in the
relevant partial waves in the 2380~MeV/$c^2$ region.
\end{abstract}

\end{frontmatter}

The experimental search for dibaryon resonances has been a long and
generally painful story, with most claims being eventually disproved.
However, the WASA collaboration working at CELSIUS has recently
produced evidence for a very significant peak in the total cross
section for the quasi-free $pn\to d\pi^0\pi^0$
reaction~\cite{BAS2009}. The experiment was carried out using a
deuterium target, with the final deuteron and the two pions from the
$pd\to pd\pi^0\pi^0$ reaction being detected and the proton in the
final state reconstructed from energy-momentum conservation. Slow
protons were then selected so that they could be treated as
\textit{spectators} that were assumed only to influence the reaction
through the kinematics. Furthermore, because the events were fully
constrained, the centre-of-mass (CM) energy $W$ of the $pn$ system
could be determined on an event-by-event basis without any
uncontrolled smearing arising from the deuteron Fermi motion. In this
way the peak region could be scanned with a proton beam of fixed
energy.

\begin{figure}[htb]
\begin{center}
\resizebox{8cm}{!}{\includegraphics[scale=1]{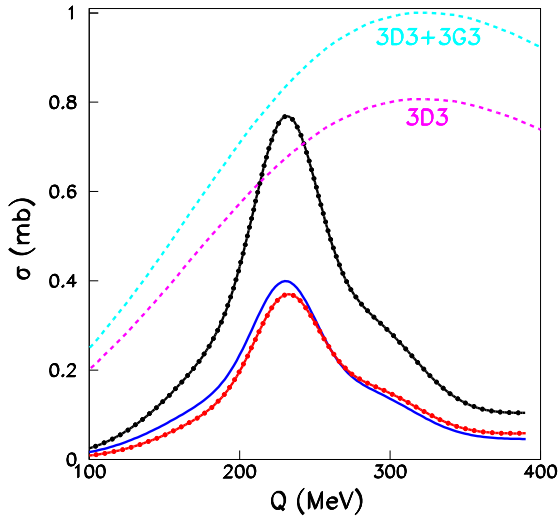}}
\caption{Total cross section for the $pn{\to}d\pi^0\pi^0$ reaction as
a function of the excess energy $Q$. The solid (blue) curve
represents a parameterisation~\cite{BAS2011} of the COSY-WASA
data~\cite{ADL2011}. The cross section predicted for
$pn{\to}pn\pi^0\pi^0$ on the basis of an $L=2$ decay using the ratio
of Fig.~\ref{ratio-plot} is shown by the dot-dashed (red) curve and
the sum of the two components by the higher dot-dashed (black) curve.
The inelastic cross sections in the $^{3\!}D_3$ (magenta) and the
$^{3\!}D_3+{}^{3\!}G_3$ (turquoise) waves taken from the SAID current
solution~\cite{SAID} and scaled by an isospin factor of $1/6$ are
shown as dashed curves. \label{money-plot} }
\end{center}
\end{figure}

After the move of the WASA facility to COSY-J\"{u}lich, the
experiment was repeated with the same basic apparatus but with
significantly higher statistical precision, which allowed much finer
divisions in the CM energy to be presented~\cite{ADL2011}. In order
to ensure that only low spectator momenta were used in the subsequent
analysis, data were in practice taken at several proton beam energies
and, although there is some uncertainty in the relative normalisation
of these runs, the good overlap lends confidence that the spectator
distribution was correctly handled when deriving cross sections. A
parameterisation~\cite{BAS2011} of the COSY results is shown in
Fig.~\ref{money-plot} in terms of the excess energy
$Q=W-(m_d+2m_{\pi^0})c^2$.

The $pn\to d\pi^0\pi^0$ total cross section is dominated by a very
strong peak at $W\approx 2.38$~GeV, \textit{i.e.}, $Q\approx
234$~MeV, and width $\Gamma\approx 70$~MeV, which the authors have
speculated to be a signal for a dibaryon resonance. Since a
$\pi^0\pi^0$ system can only have isospin $I=0$ or $I=2$, such a
resonance, as well as the dipion itself, must be isoscalar. The
likely dibaryon states would therefore be $J^p=1^+$ or $3^+$, which
couple to the initial $(^3S_1,\,^{3\!}D_1)$ or
$(^{3\!}D_3,\,^{3\!}G_3)$ proton-neutron waves, respectively.

The other important point to note is that the shape of the two-pion
spectrum from the $pn{\to}d\pi^0\pi^0$ reaction is far from that of
phase space, with an enhancement at very low $\pi^0\pi^0$ masses and
also some excess of events at maximum $m_{\pi^0\pi^0}$. These are
features that are typical of many two-pion production reactions in
the isoscalar channel and go by the name of the ABC effect from their
original observation in the $pd\to\,^3\textrm{He}(\pi\pi)^0$
reaction~\cite{ABE1960}.

The ABC effects are generally thought to be dynamical in origin,
often associated with the production of two $p$-wave pions, such as
through the excitation and decay of two $\Delta(1232)$
isobars~\cite{RIS1973}. However, it is hard to construct a model of
this type that would give a very strong peak in the overall
centre-of-mass energy, with a width that is only half that of a
single $\Delta(1232)$. The most complete evaluation in such a model
in fact gives $\Gamma\approx 200$~MeV~\cite{MOS1999}. Even though the
existence of a dibaryon might seem very improbable, the possibility
has nevertheless to be taken seriously. With that in mind, one then
has to ask if the dibaryon could and should be seen in other decay
modes. It is the purpose of the present letter to evaluate the rate
for the decay into the $pn\pi^0\pi^0$ channel in a weakly
model-dependent way by using final state interaction theory.

We have previously shown that at low proton-neutron energy $E_{pn}$
there is an intimate connection between the deuteron bound-state wave
function $\psi_{\alpha}(r)$ and that of the $pn$ $^3\!S_1$ scattering
state $\psi(q,r)$~\cite{FAL1997}. Although it is well known that the
continuation of the latter into the bound-state region does yield the
deuteron pole, it is less obvious that the residue at the pole is
uniquely fixed by the deuteron binding energy $B$. The connection is
given by
\begin{equation}
|\psi(q,r)|^2 \approx
\frac{2\pi}{\alpha(\alpha^2+q^2)}\:|\psi_{\alpha}(r)|^2 \:,
\label{theorem}
\end{equation}
where $\alpha^2=mB$, with $m$ the average nucleon mass and
$q^2=mE_{pn}$. The relation in Eq.~(\ref{theorem}) is exact at
the pole $q^2=-\alpha^2$ but at short $pn$ separations it
remains a good approximation up to energies of at least
50~MeV~\cite{FAL1997a}.

The production of a meson in nucleon-nucleon collisions necessarily
involves a large momentum transfer which is sensitive to short-range
effects in the $NN$ system. On the basis of Eq.~(\ref{theorem}) it is
easy to see that for a short-range production operator the absolute
squares of the matrix elements leading to the $dX$ and $\{pn\}_qX$
final states are related by
\begin{equation}
\left|\mathcal{M}(pn\to \{pn\}_{q}X)\right|^2 \approx \frac{2\pi m}
{\alpha(q^2+\alpha^2)} \left|\mathcal{M}(pn\to dX)\right|^2\:,
\label{elements}
\end{equation}
where $\{pn\}_q$ denotes a spin-triplet $S$-wave proton-neutron
state at relative momentum $\vec{q}$. Deviations arising from
the $pn$ tensor force and the deuteron $D$-state are also
likely to be small at short $pn$ separations.

The form of the relation in Eq.~(\ref{elements}) depends upon the
normalisation used. Here we have taken uniformly a boson
normalisation, where the relativistic $n$-body phase space is given
by
\begin{equation}
\label{phase-space} d\Phi_n(P,p_1,\cdots p_n) =
\delta^4\left(P-\sum_{i=1}^{n}p_i\right)
\prod_{i=1}^{n}\frac{d^3\vec{p}_i}{2E_i(2\pi)^3}\,,
\end{equation}
where $(E_i,\vec{p}_i)$ is the four-momentum of one of the final
particles with mass $m_i$ and $P$ is the total four-momentum. The
differential cross sections then correspond to
\begin{equation}
\label{dcs} d\sigma = \frac{1}{F}\,|\mathcal{M}|^2\,d\Phi_n\,,
\end{equation}
where the flux factor $F$ is the same for the different $n$-body
final states.

Using this approach, it is possible to describe quite accurately the
ratios of the cross sections for single-meson production in $pn\to
\{pn\}_q\eta/d\eta$~\cite{FAL2001} and $pp\to
\{pn\}_q\pi^+/d\pi^+$~\cite{FAL1997a} close to threshold. In the
following we carry out a similar analysis for two-pion production in
the WASA experiments~\cite{BAS2009,ADL2011}.

Whereas a $1^+$ dibaryon decay to $d\pi^0\pi^0$ could involve purely
$S$-waves, that of a $3^+$ state necessarily requires $L=2$ waves.
Because the observed two-pion spectrum is peaked in the $s$-wave
region of low $\pi^0\pi^0$ masses, the angular momentum must then
correspond primarily to one between the deuteron and the dipion,
where the relative momentum is $\vec{k}$. This introduces a kinematic
factor of $k^L$ in the near-threshold amplitude and this must be
included in the modelling of the cross sections. We treat separately
the cases of $L=0$ and $L=2$ in order to estimate the ratios of the
total cross sections:
\begin{equation}
\label{define} R_L(Q)=\sigma(pn\to
\{pn\}_{I=0}\pi^0\pi^0)/\sigma(pn\to d\pi^0\pi^0)\,.
\end{equation}

If the non-relativistic approximation is made to the phase space of
Eq.~(\ref{phase-space}) and one neglects any dependence of the matrix
elements upon the kinematic variables apart from the $k^L$ factor, it
is possible to obtain estimates of the cross section ratio of
Eq.~(\ref{define}) in closed form for both values of $L$:
\begin{equation}
R_{L=0}(Q) = \frac{1}{\pi\sqrt{x}}\left[\frac{8x}{15}+\frac{5}{3}
+\frac{1}{x}-\sqrt{x}\left(\frac{1+x}{x}\right)^{\!2}
\textrm{arctan}\sqrt{x}\right]. \label{ratio1}
\end{equation}
\begin{eqnarray}
\nonumber R_{L=2}(Q)&=&\frac{1}{\pi\sqrt{x}}\left[\frac{128x}{315}+
\frac{93}{35}+\frac{73}{15x}+\frac{11}{3x^2}+\frac{1}{x^3}\right. \\
&&\hspace{2cm}\left.-\sqrt{x}\left(\frac{1+x}{x}\right)^4\textrm{arctan}\sqrt{x}\right],
\label{ratio2}
\end{eqnarray}
where $x=Q/B$ and $B$ is the binding energy of the deuteron.

\begin{figure}[htb]
\begin{center}
\resizebox{8cm}{!}{\includegraphics[scale=1]{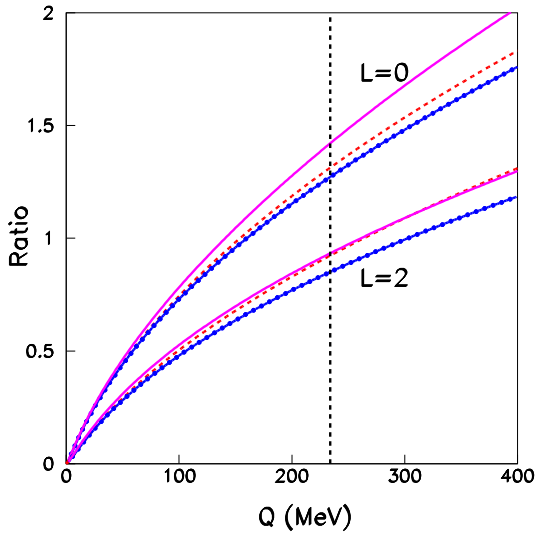}}
\caption{Predicted ratio of the
$pn{\to}pn\pi^0\pi^0/pn{\to}d\pi^0\pi^0$ total cross sections as a
function of the excess energy $Q$ for both the $L=0$ and $L=2$
possibilities. The dashed (red) curves result from the
non-relativistic analytic formulae of Eqs.~(\ref{ratio1}) and
(\ref{ratio2}), where the $\pi^0\pi^0$ spectrum is described by phase
space. These predictions are reduced slightly to the dot-dashed
(blue) curves when relativistic kinematics are used for the pions.
The introduction of an ABC-like structure in the $\pi^0\pi^0$
effective-mass distribution increases the ratio to the solid
(magenta) curves. The position of the centre of the WASA
peak~\cite{BAS2009,ADL2011} is indicated by the (black) vertical
dashed line.
 \label{ratio-plot} }
\end{center}
\end{figure}

The predictions of Eqs.~(\ref{ratio1}) and (\ref{ratio2}) are shown
in Fig.~\ref{ratio-plot}. The curve corresponding to $L=0$ lies
significantly higher than that of $L=2$, though some of this
difference might be reduced a little if barrier penetration effects
were included in the $L=2$ case~\cite{BLA1952}. For the four-body
final state some of the energy is taken by the $pn$ internal degrees
of freedom and this reduces the influence of the threshold $k^2$
factor in the $L=2$ case. The curves increase with $Q$ because of the
larger phase-space volume there for a four-body final state. It is
clear that at the indicated WASA peak position the $pn{\to}
pn\pi^0\pi^0$ and $pn{\to}d\pi^0\pi^0$ total cross sections should be
of the same order of magnitude.

It is, however, clear that for $Q$ in the 200--300~MeV range it is
dangerous to treat the pions non-relativistically, even though many
effects cancel in the cross section ratio. However, the integration
over a relativistic phase space does not lead to analytic forms
analogous to those of Eqs.~(\ref{ratio1}) and (\ref{ratio2}). By
retaining non-relativistic kinematics for the nucleons, the ratios
can be expressed in terms of two-fold integrals which have to be
evaluated numerically. This procedure reduces slightly the
predictions of the non-relativistic approach, as shown in
Fig.~\ref{ratio-plot}. The difference can be understood because, on
average, the pions in the $pn\pi^0\pi^0$ final state are slightly
less relativistic than those in $d\pi^0\pi^0$.

In the relativistic evaluation thus far presented, the two-pion
spectrum was still assumed to have a phase-space form whereas the
experimental data show a strong tendency for a peaking at low
$\pi^0\pi^0$ invariant masses~\cite{ADL2011}. This is particularly
evident at the cross section maximum, but gets softer at higher beam
energies. To investigate the effect, the $\pi^0\pi^0$ spectrum at the
``resonance'' position was fitted and it was then assumed that this
shape scaled with the available excess energy $Q$. This assumption
brings the problem closer to one of a three-body/two-body
comparison~\cite{FAL1997a,FAL2001} and increases the ratios to be
equal or even above those of the non-relativistic formulae shown in
Fig.~\ref{ratio-plot}. However, it is comforting to note that the
evaluations of the ratios in the various approximations do not give
significantly different results. The only major dependence appears to
be on the value assumed for $L$ in the decay.

\begin{figure}[htb]
\begin{center}
\resizebox{8cm}{!}{\includegraphics[scale=1]{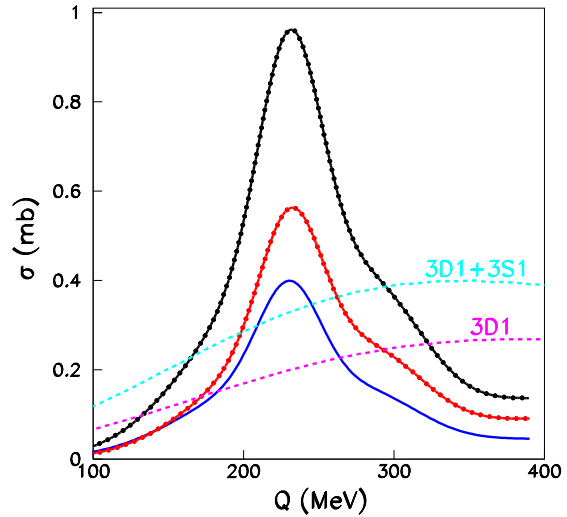}}
\caption{Analogous curves to those in Fig.~\ref{money-plot} but for
an $L=0$ decay of the dibaryon. These are compared to the inelastic
cross sections in the two $J=1$ partial waves. \label{money-plot2} }
\end{center}
\end{figure}

The predicted cross section ratio is multiplied by a parameterisation
of the WASA $pn\to d\pi^0\pi^0$ total cross section~\cite{ADL2011} in
order to get estimations for that of $pn\to pn\pi^0\pi^0$. For this
purpose no attempt was made to model the cross section peak in terms
of a resonance and background because events from the background are
likely to yield a broadly similar four-body/three-body ratio. The
results are shown separately for $L=2$ and $L=0$ in
Figs.~\ref{money-plot} and \ref{money-plot2}, respectively. It is
clear from the figures that a peak in the $pn\to d\pi^0\pi^0$ cross
section should also lead to one in $pn\to pn\pi^0\pi^0$ at a
marginally higher value of $Q$ and very slightly broadened.

Also shown in the figures are the sum of the three- and four-body
cross sections. It is then of interest to see how these compare with
the inelasticities predicted in a current partial wave analysis. It
is important to note here that the $pn\to d\pi^0\pi^0$ represents
only one sixth of a pure isoscalar $I=0$ cross section and so the
SAID predictions~\cite{SAID} for the allowed partial waves have been
reduced by a factor of six before being plotted in
Figs.~\ref{money-plot} and \ref{money-plot2}.

Although it is self-evident that the SAID predictions can only be as
reliable as the experimental input, which is rather sparse in this
energy range, the figures suggest that $L=0$ decay is not favoured
because the estimated cross section significantly exceeds the sum of
the SAID inelastic cross sections in the $^3D_1$ and $^3S_1$ partial
waves~\cite{SAID}. This point is reinforced by the fact that some of
this inelasticity will correspond to single-pion production in, for
example, $pn\to pp\pi^-$. The same argument regarding the
inconsistency with the SAID predictions cannot be made against the
$J^p=3^+$ possibility that is illustrated in Fig.~\ref{money-plot}.
This preference is consistent with the $3^+$ deduced by the
experimental group on the basis of information extracted from the
angular distributions.

Although the cross section ratio that we have evaluated has been
presented in terms of a dibaryon resonance, the work here does not
really rely on this hypothesis; it depends principally upon the
short-range nature of the two-pion production operator. The curves in
Figs.~\ref{money-plot} and \ref{money-plot2} have therefore a much
wider range of applicability and the statement regarding the strong
preference for the $J^p=3^+$ waves dominating the cross section
remains. On the other hand, if one did believe in the existence of
such a dibaryon then our work could be interpreted as finding a
relation between the $d\pi^0\pi^0$ and $pn\pi^0\pi^0$ couplings to
this state.

We have only estimated one particular inelastic channel,
\textit{viz.}\ the $pn\pi^0\pi^0$, where the proton-neutron
pair emerge in the spin-triplet $S$-wave. There are other
potential channels for the decay of a $J^p=3^+$ dibaryon, such
as $pp\pi^-\pi^0$, but their estimation falls outside the remit
of our approach and would require a full dynamical model.
Although one might question the use of the extrapolation
theorem of Eq.~(\ref{theorem}) for $Q$ in the 200-300~MeV
range, in fact the bulk of the energy is taken by the two light
particles and the fraction where the $pn$ excitation energy is
large enough to jeopardise the approach is comparatively small.
There are probably greater uncertainties associated with the
assumptions of the dependence of the matrix element upon the
possible kinematic variables. However the strong distortion of
the $\pi^0\pi^0$ invariant mass spectrum only changes
predictions by typically 10\%. The angular momentum in the
decay seems to make the much bigger difference illustrated in
Fig.~\ref{ratio-plot}.

\newpage
Independent of the origin of the peak in the $pn\to d\pi^0\pi^0$
total cross section identified by the WASA
collaboration~\cite{BAS2009,ADL2011}, the approach proposed here
shows that there should be an analogous peaking in $pn\to
pn\pi^0\pi^0$ cross section of a similar strength. This channel could
be investigated using the existing WASA data, though the kinematics
of this state are harder to identify due to the presence of the final
neutron~\cite{Heinz}.

\vspace{5mm}

The inspiration for this work came from many discussions with members
of the WASA collaboration, in particular with M.~Bashkanov,
H.~Clement, and C.~Hanhart, for which we are very grateful. The
parameterisation of the COSY-WASA data used here was kindly provided
by M.~Bashkanov. Correspondence with D.~V.~Bugg has also proved very
useful. This work was supported by the European Community under the
``Structuring the European Research Area'' Specific Programme
Research Infrastructures Action (Hadron Physics, contact number
RII3-cT-204-506078), and by the Swedish Research Council.

%
%

%
%
\end{document}